\documentclass[]{raa}

\usepackage{graphicx,times}
\input{epsf.sty}
\input{psfig.sty}

\begin{document}

   \title{Investigating the Disk-Corona Relation in a Blue AGN Sample
}

\volnopage{ {\bf 2009} Vol.\ {\bf X} No. {\bf XX}, 000--000}
   
   \setcounter{page}{1}

\author{Jie-Ying Liu
      \inst{1,2}
   \and B. F. Liu
      \inst{1}
         }

   \institute{National Astronomical Observatories/Yunnan Observatory, Chinese Academy of Sciences,
             PO Box 110, Kunming 650011; $ljy0807@ynao.ac.cn$\\
        \and
             Graduate School of Chinese Academy of Sciences, Beijing 100049 \\
\vs \no
   {\small Received [year] [month] [day]; accepted [year] [month] [day] }
}
   \abstract{
We compile a blue AGN sample from SDSS and investigate the ratio of hard X-ray
to  bolometric luminosity in dependence on 
Eddington ratio and black hole mass.  Our sample comprises 240 radio-quiet  Seyfert 1 galaxies and QSOs. We find
that the fraction of hard X-ray luminosity (log$(L_{\rm 2-10
kev}/L_{\rm bol})$) decreases with the increase of Eddington ratio. 
We also find that
the fraction of hard X-ray luminosity is independent on the black
hole mass for the radio-quiet AGNs. The relation of log$(L_{\rm 2-10
kev}/L_{\rm bol})$ decreasing with increasing Eddington ratio indicates that 
X-ray bolometric correction is not a constant, from a larger sample supporting  the results of Vasudevan \& Fabian (2007).  We
interpret our results by the disk corona evaporation/condensation
model (Meyer et al.~\cite{me200}; Liu et al. 2002a; Liu et al.~2007). In the
frame of this model, the Compton cooling becomes efficient in
cooling of the corona at high accretion rate (in units of Eddington rate), leading to
condensation of corona gas to the disk. Consequently, the relative
strength of corona to the disk becomes weaker at higher Eddington
ratio. Therefore, the fraction of hard X-ray emission to disk emission and hence to the bolometric emission is smaller at higher Eddington ratio. The independence of the fraction of hard X-ray luminosity on the
mass of the black hole can also be explained by the disk corona model
since the corona structure and luminosity (in units of Eddington
luminosity) are independent on the mass of black holes. 
   }

   \authorrunning{J. Y. Liu \& B. F. Liu }            
   \titlerunning{Disk-Corona Relation in Blue AGNs}  

   \maketitle

\section{Introduction}
\label{sect:intro}

The accretion onto the supermassive black hole is the key process to produce
the  spectral energy distribution in AGNs, including the Big Blue Bump,
soft X-ray excess, Fe $K\alpha$ lines at 6.4 keV and hard X-ray
tail. These observed  spectral properties indicate that the cold and hot gas
coexist near the central supermassive black hole. According to the
 commonly accepted theory the optical/UV radiations are emitted from a geometrically
thin and optically thick disk (i.e. the cold gas) (e.g., Shakura \&
Sunyaev~\cite{SS73}, Lynden--Bell \& Pringle~\cite{lp74}), and the hard
X-rays arise from hot optically thin and geometrically thick
accretion flows, such as disk corona (Haardt
\& Maraschi 1991, 1993; Nakamura \& Osaki 1993; Svensson \& Zdziarski 1994; Kawaguchi et al. 2001, Cao 2009) or advection--dominated accretion flows (ADAF)(e. g., Narayan \& Yi~\cite{ny94}, 1995; Narayan et al. ~\cite{na98}).

The disk corona model was proposed by Liang \& Price (\cite{lp77}) to
explain the observed X-ray emission in Cyg X--1. The accretion in the corona is supplied with material and energy through disk evaporation 
(e.g., Meyer \& Meyer--Hofmeister~\cite{mm94}; Liu et
al.~\cite{liu99}) or
 magnetic process (e.g. Galeev et al.~\cite{gal79}; Stella et al.~\cite{stell84}; Liu et al.2002b,~\cite{liu03}). In AGNs a hot corona sandwiched by a disk is thought responsible to emit the X-rays (e.g. Haardt \& Maraschi 1991; Nakamura \& Osaki 1993; Kawaguchi et al. 2001; Liu et al. 2002b, 2003). In the context of these studies it is often  assumed that a fraction ($f$) of accretion energy is released in the corona in fits to individual AGN spectrum.   Questions arise as, is the fraction $f$ dependent on the Eddington ratio and black hole? how does the relative strength of the corona to the disk vary with the Eddington ratio and black hole mass? 

In this work, we attempt to investigate these issues from observation.
We compile a large blue AGN sample of 240 objects
derived from Dong et al.~\cite{dong08} (hereafter Dong08), in which the objects are selected from the Sloan Digital Sky
Survey (SDSS) DR4 spectral data and less affected by dust extinction. We
investigate the  correlation between the corona component and Eddington ratio. We show how the coronal luminosity fraction varies with the Eddington ratio and black hole mass for radio-quiet (RQ)
AGNs. The results  are compared with previous results of Wang et
al. (\cite{wang04}) for a sample of 56 RQ AGNs observed by ASCA and Yang et al. (\cite{yang07}) for
a combined sample from different observational satellites.
Our aim is to study the physical relation between the disk and corona. 

 The outline of the paper is in the following.  In Sect.2  the black hole mass, the optical and X-ray luminosities are calculated.  In Sect.3 the overall properties of the sample and the correlation analyses are present.  In the Sect.4 the statistical results are interpreted by the disk corona model. The discussion and conclusions are given in Sect.5 and Sect.6 respectively.

\section{The Sample and Data Reduction}
\label{sect:sample} Our sample is derived from a blue AGN sample in
Dong08. These blue AGNs are selected from the Sloan Digital Sky
Survey (SDSS) DR4 spectral data, including Seyfert 1 galaxies and
QSOs. They are less affected by dust extinction. In order to
investigate the relation between the disk and the corona, whether and how
the black hole mass and Eddington ratio affect this relation, we
select the objects by the following criteria:

\begin{itemize}
\item (1) The objects are detected or covered by Faint Images of Radio Sky
 at Twenty Centimeters (FIRST)
 survey. 
\item (2) The X-ray flux is available.
\item (3) Combining the radio with the optical data, we can determine the radio loudness and the RQ AGNs are selected.

\end{itemize}

 According to the above criteria, we analyze the sample in Dong08.
 Given the upper limit to the radio flux, 404 AGNs are detected or covered by
  FIRST. After matching to ROSAT bright and faint SRC Catalogue
database ( ROSAT--SRC) with a distance criterion between the optical
and X-ray source of $\Delta \leq 60''$ (Voges et
al.~\cite{voges94}), we obtain a sample containing 258 AGNs.

  We use the radio--loudness definition (Ivezi\'{c} et al.~\cite{Ive02}) to
   divide RQ and radio-loud(RL) AGNs, that is,
\begin{equation}
R_{\rm i} \equiv \log(\frac{F_{\rm 20cm}}{F_{\rm i}})=0.4(i-t).
\end{equation}
where $F_{\rm i}$ and $F_{\rm 20cm}$ are flux densities at I--band
and 20cm respectively, $i$ is the I--band magnitude, and
$t=-2.5\log(F_{\rm 20cm}/3631 \rm Jy)$. Objects with $R_{\rm i}> 1$
are RL and objects with  $R_{\rm i}< 1$ are RQ. According to this
definition, 240 (93\%) objects in the sample are RQ and 18 (7\%)
objects are RL. We mainly analyze these 240 RQ objects in order to constraint the accretion mechanism.

  The black hole mass, $M_{\rm BH}$, is a fundamental parameter of AGNs.
  Different methods have been developed to estimate the black hole mass
   (Woo \& Urry~\cite{woo02} and reference
therein; Vestergaard \& Peterson~\cite{vest06}),
of which the reverberation mapping is the most widely used method 
(Peterson ~\cite{peterson93}; Peterson et al.~\cite{peterson04}).
 In the reverberation mapping method, the distance of the broad line region (BLR) from the
central black hole can be  deduced from the time lag between
continuum and emission lines. By combining the distance with the mesaured emission line width, the black hole mass can be determined.
 In our sample, we take the
$FWHM(H\beta)$ as the circular velocity in BLR. The size/distance of the BLR is given by an empirical relation with the continuum luminosity at $5100\AA$ (Vestergaard \& Peterson 2006).
 Thus, the mass of black hole for RQ AGNs
can be estimated as

\begin{equation}
\log M_{\rm BH}(H\beta)=\log\left\{\left[\frac{FWHM(H\beta)}{1000~{\rm km~s^{-1}}}\right]^2\left[\frac{\lambda~L_{\lambda}({\rm 5100\AA})}{10^{44}~{\rm ergs~s^{-1}}}\right]^{0.5}\right\}+(6.9\pm0.02).
\end{equation}

 The bolometric luminosity of
  AGNs is approximately calculated from optical continuum luminosity,
  $L_{\rm bol}\approx 9\lambda~L_{\lambda}({\rm 5100\AA})$
  (Elvis et al.~\cite{elvis94}). The Eddington ratio ($ L_{\rm bol}/L_{\rm Edd}$) is then calculated from the bolometric luminosity and black hole mass,
\begin{equation}
 L_{\rm bol}/L_{\rm Edd}=0.072{\lambda~L_{\lambda}({\rm 5100\AA})\over 10^{44}{\rm ergs~s^{-1}}}\left(M_{\rm BH}\over 10^8M_\odot\right)^{-1},
\end{equation}
where the Eddington luminosity is  $L_{\rm Edd}=1.25\times 10^{38}M_{\rm BH}/M_{\odot} ~\rm ergs~s^{-1}$.

We get the X-ray flux densities from the count rates using
    the energy to counts conversion factor (ECF) for power-law spectra and
    Galactic absorption, where the power-law photon indices
    and the corresponding absorption column densities ($N_H$) are estimated
    from the two hardness ratios given by the ROSAT-SRC. With the integrated flux
    densities in 0.1--2.4 keV and power-law indices $\Gamma$ we deduce the monochromatic flux at 2keV.
After deriving the flux at 2 keV we extrapolate
 X-ray flux ($F_{\rm int}$) in the range from 2 keV to 10 keV by assuming  a constant
 hard X-ray photon index, $\Gamma=1.9$ (Pounds et al.~\cite{pounds90}; Reeves \& Turner~\cite{retu00}).
  The X-ray luminosity is then calculated by the standard luminosity--flux relation,
\begin{equation}                                                      
L=4\pi d_{\rm L}^2F_{\rm int}(1+z)^{-(1-\alpha)},
\end{equation}
where $\alpha=\Gamma-1$ and $d_{\rm L}$ is the luminosity distance
related to the redshift and  cosmology model. 

  Throughout this paper, we adopt the cosmology model with $H_0=70{\rm km~s^{-1}Mpc^{-1}}$,
  $\Omega_{\rm M}=0.3$, $\Omega_{\Lambda}=0.7$.

 Our sample is listed in Table 1 of Appendix A, where column 1 is the object name,
 column 2 the redshift, column 3 the derived mass of black hole, column 4 the optical
luminosity at 5100\AA (in unit of $10^{44}$~ergs~s$^{-1}$), column 5 the radio flux at 20cm, column 6 the
radio--loudness, column 7 and 8 are the derived X-ray data in
0.1--2.4 keV and 2--10 keV respectively.

\section{Data Analysis and Results}
\label{sect:analysis}
\subsection{Properties of the Sample}
\begin{figure}
  \vspace{2mm}
\begin{center}
 \hspace{3mm}
\psfig{figure=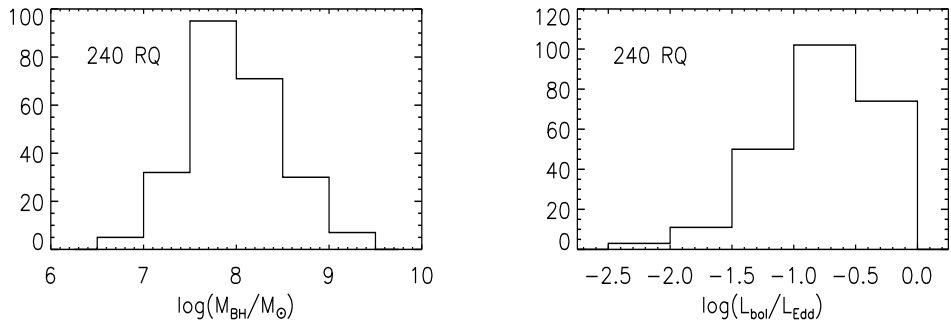,width=140mm,height=50mm,angle=0.0}
\caption{  The left plot is the histogram of black hole mass for our sample.
 We can see that the center of the distribution of $M_{\rm BH}$
of the RQ sources is nearly $10^{8.0} M_{\odot}$. The right panel plots the histogram of Eddington ratio for the sources. Most of sources are at high accretion rate. }
   \label{Fig:histograms}
  \end{center}
\end{figure}

  Fig.1 shows the histogram of black hole mass and the Eddington ratio of
  our sample. It can be seen that the mass of black holes covers a large
  range from $10^{6}$ to $10^{10}M_{\odot}$. The center of the distribution of $M_{\rm BH}$ for the sources is
greater than $10^{7} M_{\odot}$. The histogram of
the Eddington ratio of the sample shows that the Eddington ratio ranges
from $10^{-2.5}$ to 1.0, most of the objects are at high accretion
rate. If the accretion mode in AGNs is similar to that in black
hole X-ray binaries, the high accretion rates of our sample indicate
that the thin disk extends down to the innermost stable circular
orbit, hence the hard X-ray arises from a disk corona rather than an ADAF.

\subsection{Correlation Analysis and Results}

  We plot the ratio of X-ray luminosity to the bolometric luminosity
  vs. the mass of black holes and vs. the Eddington ratio for RQ sources
  in Fig.2. In the upper two panels the vertical axis is
logarithm of the fraction of hard X-ray luminosity to bolometric
luminosity (log($L_{\rm 2-10 keV}/L_{\rm bol}$)). While in the lower
panels, the vertical axis is logarithm of the fraction of soft X-ray
luminosity (0.1--2.4 keV) to the bolometric luminosity. Comparing
the upper and lower panels one can see that the distribution trends
are different for soft X-ray and hard X-ray fraction. 
\begin{figure}
   \vspace{2mm}
   \begin{center}
   \hspace{3mm}\psfig{figure=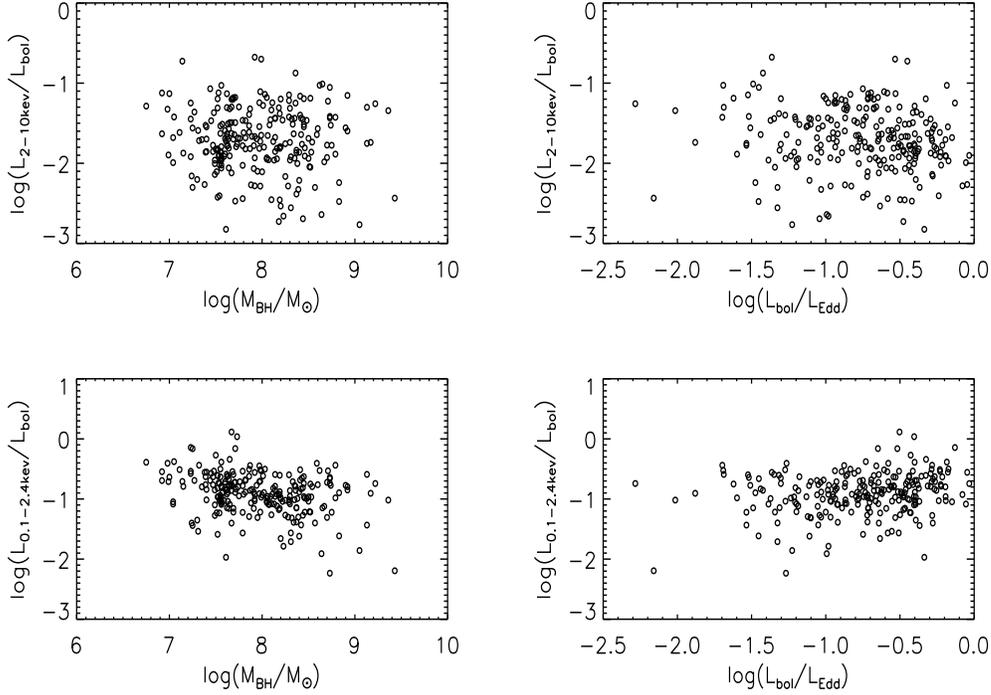,width=140mm,height=100mm,angle=0.0}
\caption{The fraction of X-ray luminosity in bolometric luminosity
vs. the mass of black hole and Eddington ratio for our sample. The top panels show log($L_{\rm 2-10 keV}/L_{\rm
bol}$) in dependence on log$M_{BH}$ and log$(L_{\rm bol}/L_{\rm Edd})$. The bottom
panels describe the trend of log($L_{\rm 0.1-2.4 keV}/L_{\rm bol}$)
varying with log$M_{\rm BH}$ and log$(L_{\rm bol}/L_{\rm Edd})$. We can see that
log($L_{\rm 0.1-2.4 keV}/L_{\rm bol}$) increases with the decrease
of black hole mass and the increase of the Eddington ratio. }
\label{Fig:240 RQ AGNs}
   \end{center}
\end{figure}




We test the correlations between log($L_{\rm 2-10 keV}/L_{\rm bol}$) and 
log($M_{\rm BH}$) and between log($L_{\rm 2-10 keV}/L_{\rm bol}$) and Eddington ratio by the method of
Spearman's rank correlation analysis (William et
al.~\cite{will00}). We calculate the Spearman's rank correlation
coefficient($\rho$) and the two-sided significance of its deviation
from zero ($P_{\rm Null}$), representing the probability for
randomness of the observed correlation of two variables. Significant
correlation of two variables is accepted if $P_{\rm null}$ is less
than 0.05. For our sample, the coefficient $\rho$ between log($L_{\rm 2-10
keV}/L_{\rm bol}$) and log$M_{\rm BH}$ is 0.001, and $P_{\rm Null}$
is 0.988. This means that the fraction of hard X-ray  is independent
of black hole mass. For the relation between the hard X-ray
fraction and Eddington ratio, we get $\rho=-0.17$, $P_{\rm
Null}=0.008$. This indicates that for RQ AGNs the hard X-ray fraction correlates
with Eddington ratio, with increase of the Eddington ratio, the hard
X-ray fraction decreases. This result is qualitatively consistent with the work of
Wang et al. (2004). 

In addition, we analyze  the correlation of  soft X-ray
 fraction with the black hole mass and Eddington ratio. It is found that the soft
  X-ray fraction in the bolometric luminosity depends on black hole mass
 ($\rho=-0.26$, $P_{\rm Null}=4.085e-5$), in contrast to the case
  of hard X-rays. The soft X-ray fraction also  positively correlated with the Eddington ratio ($\rho=0.18$,
 $P_{\rm Null}=0.005$),
 contrary to the trend of hard X-ray fraction. 



\section{Interpretation of the Correlations with Disk Corona Model}

The hard X-ray radiation from RQ AGNs with high Eddington ratio is
commonly thought to be produced from a disk corona as a result of
Comptonization of the softer photons arising from the accretion
disk (e.g. Haardt \& Maraschi 1993; Kawaguchi et al. 2001; Liu et al. 2003; Cao 2009). If the bolometric luminosity is contributed dominantly by the
disk ($L_d$) and corona ($L_c$),  the fraction of hard X-ray
luminosity in the total luminosity represents the strength of the
corona relative to the disk. Thus, the negative correlation between
the fraction of hard X-ray and Eddington ratio for RQ AGNs indicates
that the corona relative to the disk becomes weaker as the Eddington-scaled accretion rate increases. 

Theoretical calculation of radiations from the disk corona  (Haardt \& Maraschi 1991; 1993) shows that the spectral shape does not change with accretion rate. In other words, ratio ($f$) of X-ray luminosity and bolometric luminosity is predicted to be independent on the accretion rate. To explain the observational correlation  between $f$ and Eddintong accretion rate, Wang et al. (2004), Yang et al. (2007) and Cao(2009) consider magnetic fields and different shear stress tensors. 
Here we introduce the disk corona evaporation and/or condensation model (Liu et al. 2002a). We show that even without magnetic fields the correlation can be simply a consequence of efficient condensation of coronal gas to the disk in the case of high accretion rates.

Assuming that the energy of photons from the disk is amplified by
 a factor of $A\equiv \Delta \epsilon /\epsilon $ in the process of collisions with electrons when  going
  through the corona, for non-relativistic thermal distributions of electrons $A$ is a function of electron temperature ($T_e$) and optical
depth ($\tau_{es}$) of the corona, 
\begin{equation}
A={4kT_e\over m_ec^2}\tau_{es}.
\end{equation}
The fraction of hard X-ray luminosity  is expressed as,
\begin{equation}\label{X-farction}
{L_{2-10keV}\over L_{\rm bol}}={AL_d\over AL_d+L_d}={1\over 1+{1\over A}}.
\end{equation}
In the case of high accretion rate, disk radiations are strong, which leads to efficient inverse Compton scattering in the corona. As a consequence of over-cooling, part of the corona gas condenses into the disk, thereby the optical depth decreases and hence the amplification factor is low. The higher the accretion rate, the lower density of the corona is. Therefore, from Eq.\ref{X-farction} one can see that the fraction of X-ray decreases at high accretion rates.
Thus, the negative correlation between the fraction of hard
X-ray luminosity and the Eddington ratio can be explained by the corona condensation  to the disk as a consequence of pressure and energy balance between the disk and corona.

Detailed  calculations of condensation rate and luminosity have been carried out by Liu et al. (2007) and Taam et al. (2008). In this model, when the Compton scattering is the dominant cooling mechanism, the luminosities from the corona and disk are respectively as follows (Taam et al. 2008),
\begin{eqnarray}\label{lcom}
L_{\rm c}&=&\int_{3R_{s}}^{R_{out}}2\pi RH\frac{4kT_{\rm e}}{m_{\rm e}c^2}n_{\rm e}\sigma_{\rm T}4\sigma T^4_{\rm eff}(R)dR \nonumber\\
&=&4.11\times 10^{38}\alpha^{-\frac{7}{5}}m\dot{m_{\rm c}}^{\frac{7}{5}}\dot{m_{\rm d}}^{\frac{3}{5}}\int_{3}^{r_{out}}\left(\frac{3}{r}\right)^{\frac{23}{10}}\left[1-\left(\frac{3}{r}\right)^{\frac{1}{2}}\right]^{\frac{3}{5}}dr.
\end{eqnarray}
\begin{eqnarray}\label{ldisk}
L_{\rm d}&=&\int_{3R_s}^{R_{out}}\sigma T^4_{\rm eff}(R)4\pi RdR \nonumber\\
&=&\int_{3R_s}^{R_{out}}\frac{3GM\dot{M_{\rm d}}}{8\pi R^3}\left[1-\left(\frac{3R_s}{R}\right)^{\frac{1}{2}}\right]4\pi RdR \nonumber\\
&=&9.38\times 10^{38}m\dot{m_{\rm d}}\int_{3}^{r_{out}}\frac{1}{r^2}\left[1-\left(\frac{3}{r}\right)^{\frac{1}{2}}\right]dr.
\end{eqnarray}
where $r=R/R_{s}$ ($R_{s}$ the Scwarzschild radius), $m=M_{\rm BH}/M_\odot$, $\dot m_c$ is the Eddington-scaled accretion rates in the corona and $\dot m_d$ the one in the disk. Note that for a truncated inner disk lying under a corona, $\dot m_d$ comes only from the condensation of coronal gas; While in the case of a full disk coexisting with a corona for AGNs discussed here, the accretion rate in the disk is mainly from outer region, $\dot m_d=\dot m-\dot m_c$.    
We then derive the ratio of corona luminosity and disk luminosity,
\begin{equation}\label{lc/ld}
 L_{\rm c}/L_{\rm d}\propto \alpha^{-\frac{7}{5}}\dot{m_{\rm c}}^{\frac{7}{5}}\dot{m_{\rm d}}^{-\frac{2}{5}}=\alpha^{-\frac{7}{5}}\dot{m_{\rm c}}^{\frac{7}{5}}(\dot m-\dot{m_{\rm c}})^{-\frac{2}{5}}.
\end{equation}
In the disk corona evaporation and/or condensation model, the coronal accretion flow is supplied by disk evaporation in the region around $300 R_s$. The coronal flow rate $\dot m_c$ can be as high as a few percent of Eddington value if the accretion rate exceeds a few percent of Eddington value. However,  the coronal flow can condense partially into the disk in the inner region as a result of  Compton cooling, and the condensation rate is larger at higher accretion rate. This means that the coronal flow ($\dot m_c$) is lower at higher accretion rate.  From Eq.(\ref{lc/ld}) one can see that the ratio of corona luminosity and disk luminosity decreases with increasing accretion rate $\dot m$. Noting that the accretion rate in unit of Eddington value equals to the Eddington ratio, the disk corona evaporation and condensation model predicts that the ratio of X-ray luminosity and disk luminosity decreases with increasing Eddington ratio, which can explain the correlation obtained from our RQ AGN sample.

If magnetic fields are also taken into account as additional heating (Liu et al.~2003; Cao 2009 ), the coronal accretion flow can be higher and  the corona can be stronger than that without magnetic fields. But the trend that condensation is stronger at higher accretion rates does not change by the magnetic fields. Therefore, the variation of relative strength of the disk and corona with accretion rates keeps the same trend. Thus, with inclusion of magnetic fields the model will still work in interpreting the correlations.    

Another important feature of the disk corona evaporation and condensation model is that the corona structure and evaporation/condensation properties are independent on the mass of black hole if scaled properly. In fact, the model can be applied in both X-ray binaries and AGNs. Eq.(\ref{lc/ld}) shows that the luminosity ratio from the corona and disk indeed does not vary with the mass of black hole. Therefore, the model can explain why for our sample the 
 fraction of hard X-ray luminosity is independent on
the mass of black hole.

\section{Discussion}
\subsection{Implication to the Hard X-ray Bolometric Correction}
The bolometric luminosity is either estimated from the measured  luminosity at $5100\AA$ or from hard X-ray luminosity with fixed correction factors. However, our data analysis shows that the hard X-ray fraction in the bolometric luminosity decreases with increasing Eddington ratio for RQ AGNs.  This means that bolometric correction factor from hard X-rays,  $k_{\rm 2-10keV}=L_{\rm bol}/L_{2-10keV}$, should increase with increasing Eddington ratio. This result is in good agreement with that of Vasudevan \& Fabian (2007), which is drawn from 54 AGNs with available SEDs. Note that in our study we calculate the bolometric luminosity from $9\lambda L_\lambda(5100\AA)$. If the bolometric correction factor for $L_{5100\AA}$ also tends to increase with Eddington ratio (see Fig.15 of Vasudevan \& Fabian 2007), the hard X-ray fraction in bolometric luminosity would be even smaller at higher Eddington ratio for our sample. Therefore, our statistical results support that the bolometric correction for hard X-rays increases with increasing Eddington ratio.

\subsection{The Constant Photon Index}
In order to investigate the strength of corona ie. to get the hard X-ray flux, we use the same photon indices for all objects in our sample to extrapolate the hard X-ray emissiong from 2keV flux. From above analysis, we get a smaller absolute coefficient value ($\rho$) between the fraction of hard X-ray luminosity and Eddington ratio. However, if we
note that for some individual objects the spectra become softer when a source is brighter (e.g.,
Lu \& Yu~\cite{luyu99}, Bian et al.~\cite{bian05}, Middleton et
al.~\cite{mid08}; Saez et al. 2008) and this effect is taken into account, the hard X-ray luminosity fraction should change with the
Eddington ratio more steeply than our results  since here we assume
the photon index is constant in calculating the hard X-ray
luminosity.
\subsection{The Origin of Soft X-rays in 0.1-2.4 keV }
Our study also indicates that for the same RQ AGNs  the
fraction of soft X-ray luminosity in the ROSAT energy
band ($0.1-2.4 \rm keV$) anticorrelates on the mass of black hole, as
shown in Fig.2.
If the soft X-ray is also from the corona,
the correlation should be similar to that for hard X-rays, that is,
the fraction of soft X-ray is independent on the balck hole mass.
On the other hand, if the disk
emits  dominantly in the soft X-ray, the fraction of the soft X-ray
to the total luminosity should also be independent on the mass of
black hole since both the total luminosity and disk luminosity are
proportional to the mass of black hole. We note  that the effective
temperature of the disk decreases with the mass of black hole (e.g.,
Thorne K. S.~\cite{thone74}, Bonning et al.~\cite{bon07}),
$\sigma T^4_{\rm eff}={3GM\dot M\over 8\pi R^3}\left[1-(3R_s/R)^{1/2}\right]$,
reaching a maximum at $R=(49/36)(3R_S)$ (see Liu et al. 2007),
\begin{equation}
T_{\rm eff,max}=1.3\times 10^5 (M_{\rm BH}/10^8M_\odot)^{-1/4}{\dot m}^{1/4}K.
\end{equation}
For AGN with a high mass of black hole, the disk does not emit in
the soft X-rays but in the optical/UV band.  With mass decreases,
the high energy part of the disk spectrum could extend to soft
X-ray. Consequently, the fraction of disk contribution to the soft
X-ray increases with decreasing of $M_{\rm BH}$. This could lead to a
negative correlation between the fraction of soft X-ray and the mass
of black hole, qualitatively consistent with the observational
results  shown in Section 3. Therefore,  the different
correlations with black hole mass of the soft X-ray
and hard X-ray may indicate that the
soft X-rays are contributed from not only the corona but also the disk.
Zhou et al. (1997) separated two components in the soft X-ray region through statistic analyzing the Big Blue Bump.
 
\section{Conclusion}
\label{sect:conclusion}
  We compile a blue AGNs sample containing Seyfert 1 galaxies and QSOs with 240 AGNs
being RQ. We mainly investigate the dependence of the
 fraction of hard X-ray luminosity in bolometric luminosity
(log$(L_{\rm 2-10 keV}/L_{\rm bol})$) on the black hole mass and
 Eddington ratio in RQ AGNs by using the Spearman's rank
correlation. We find that the fraction of hard X-ray luminosity in
bolometric luminosity decreases with the increase of Eddington ratio
for RQ AGNs. While the fraction of hard X-ray luminosity in the
bolometric luminosity is independent on black hole mass for a wide
range of black hole mass. The correlation between the hard X-ray and bolometric luminosity suggests that the hard X-ray bolometric correction increase at high Eddington ratio, confirming with a large sample  Vasudevan \& Fabian (2007)'s results.  
These observational features support
the disk corona model developed by Meyer et al. (2000) and Liu et al. (2002a).
Combining the
theoretical model and the wide range of black hole mass in our
sample, we suggest that the intrinsic accretion process in AGNs and
black hole X-ray binaries be similar.

\normalem
\begin{acknowledgements}
  We are grateful to the anonymous referee for constructive and enlightening comments. Liu J. Y. greatly thanks Yuan W. for the instruction in X-ray data
deducing and the statistical analysis. We also acknowledge Dong X. B. for
the SDSS data analysis. This work is supported by the
National Natural Science Foundation of China (grants 10533050 and
10773028) and National Basic Research Program of China-973 Program 2009CB824800.
\end{acknowledgements}

\clearpage
\appendix
\section{Table}
\begin{tiny}
\begin{center}
\begin{tabular}{ccccrrcc}
\multicolumn{8}{c}{\normalsize{Table1.The Blue AGNs Sample}}\label{sample}\\
\hline
\hline
\textbf{Object}&\textbf{Z}&\textbf{log($M_{\rm BH}$)}&\textbf{$L_{\rm 5100\AA}$}&\textbf{$F_{\rm 20cm}$}&\textbf{$R_{\rm i}$}&\textbf{log($L_{\rm 0.1-2.4keV}$)}&\textbf{log($L_{\rm 2-10keV}$)} \\
\textbf{SDSS}& & ($\rm M_{\odot}$)& ($10^{44}$~ergs~s$^{-1}$)  & (mJy) & & ($\rm ergs~s^{-1}$) & ($\rm ergs~s^{-1}$) \\
\textbf{(1)}&\textbf{(2)}&\textbf{(3)}&\textbf{(4)}&\textbf{(5)}&\textbf{(6)}&\textbf{(7)}&\textbf{(8)} \\
\hline
000710.01$+$005329.0 & 0.31620 & 9.13 & 3.85 &  1.44    &  0.55 &  44.949 & 44.238 \\
000834.72$+$003156.1 & 0.26303 & 7.71 & 2.71 &  $<$1.00 &  0.43 &  44.665 & 42.915 \\
000943.14$-$090839.1 & 0.20958 & 8.52 & 2.09 &  $<$1.00 &  0.21 &  44.057 & 43.225 \\
001247.93$-$084700.4 & 0.22006 & 8.01 & 1.95 &  $<$1.00 &  0.21 &  44.629 & 43.965 \\
004319.74$+$005115.3 & 0.30807 & 9.36 & 3.08 &  1.60    &  0.80 &  44.425 & 44.099 \\
010226.31$-$003904.6 & 0.29421 & 7.92 & 9.63 &  $<$1.00 &$-$0.10&  45.012 & 43.656 \\
011254.92$+$000313.0 & 0.23855 & 8.14 & 1.98 &  $<$1.00 &  0.42 &  44.398 & 43.816 \\
014017.07$-$005003.0 & 0.33424 & 8.92 &13.40 &  $<$1.00 &  0.06 &  45.232 & 44.927 \\
014234.40$-$011417.3 & 0.24453 & 7.14 & 0.68 &  $<$1.00 &  0.69 &  44.078 & 44.060 \\
015530.01$-$085704.0 & 0.16465 & 8.48 & 1.78 &  $<$1.00 &$-$0.04&  44.212 & 43.739 \\
015910.04$+$010514.6 & 0.21716 & 8.08 & 1.99 &  $<$1.00 &  0.22 &  44.164 & 43.731 \\
022347.48$-$083655.5 & 0.26077 & 7.52 & 0.97 &  1.04    &  0.81 &  44.330 & 43.631 \\
022417.16$-$092549.3 & 0.31148 & 8.79 & 1.72 &  $<$1.00 &  0.67 &  44.751 & 43.763 \\
072937.04$+$375435.0 & 0.20364 & 7.23 & 1.39 &  $<$1.00 &  0.40 &  44.558 & 43.179 \\
073503.49$+$431153.5 & 0.26246 & 8.11 & 1.86 &  $<$1.00 &  0.59 &  44.302 & 43.979 \\
074645.05$+$314149.2 & 0.32696 & 7.55 & 3.31 &  $<$1.00 &  0.54 &  44.972 & 43.504 \\
074820.97$+$340752.6 & 0.34304 & 8.09 & 3.08 &  $<$1.00 &  0.68 &  44.482 & 43.986 \\
074948.26$+$345444.0 & 0.13181 & 7.71 & 1.16 &  0.82    &$-$0.01&  44.129 & 43.526 \\
075217.84$+$193542.2 & 0.11723 & 9.43 & 2.59 & 22.95    &  0.84 &  43.173 & 42.932 \\
075245.60$+$261735.7 & 0.08216 & 6.92 & 0.46 &  1.27    &  0.13 &  43.924 & 42.985 \\
075819.68$+$421935.1 & 0.21125 & 7.97 & 2.84 &  1.81    &  0.36 &  44.958 & 43.865 \\
075949.54$+$320023.9 & 0.18805 & 7.60 & 1.92 &  $<$1.00 &  0.35 &  44.488 & 43.353 \\
080559.94$+$260602.3 & 0.13594 & 8.31 & 1.34 &  $<$1.00 &$-$0.09&  43.371 & 42.525 \\
080644.65$+$384318.3 & 0.34539 & 7.98 & 3.28 &  $<$1.00 &  0.71 &  44.954 & 44.363 \\
081054.73$+$501319.5 & 0.32184 & 9.13 & 5.47 &  $<$1.00 &  0.37 &  44.258 & 43.940 \\
081317.91$+$435620.7 & 0.25455 & 8.30 & 1.81 &  $<$1.00 &  0.60 &  44.351 & 43.472 \\
081427.69$+$433705.1 & 0.22418 & 7.61 & 1.27 &  $<$1.00 &  0.47 &  44.205 & 43.334 \\
081738.33$+$242330.0 & 0.28259 & 8.20 & 4.23 &  4.45    &  0.86 &  44.599 & 44.382 \\
082633.51$+$074248.4 & 0.31064 & 8.14 & 7.54 &  $<$1.00 &  0.12 &  44.876 & 43.978 \\
082640.73$+$063041.5 & 0.17092 & 7.82 & 0.80 &  $<$1.00 &  0.50 &  44.165 & 43.175 \\
083120.99$+$483154.4 & 0.33466 & 7.56 & 3.30 &  $<$1.00 &  0.62 &  45.087 & 44.444 \\
083225.34$+$370736.1 & 0.09191 & 9.22 & 1.20 & 11.73    &  0.75 &  44.291 & 43.776 \\
083417.91$+$491439.2 & 0.17315 & 7.40 & 0.93 &  $<$1.00 &  0.26 &  44.082 & 43.225 \\
083443.80$+$382632.7 & 0.28822 & 8.59 & 2.94 &  $<$1.00 &  0.48 &  45.016 & 43.780 \\
083553.46$+$055317.1 & 0.20438 & 7.49 & 2.24 &  $<$1.00 &  0.15 &  44.031 & 43.165 \\
083658.91$+$442602.4 & 0.25443 & 8.74 &13.58 & 10.25    &  0.64 &  45.314 & 44.562 \\
084230.51$+$495802.3 & 0.30508 & 8.53 & 2.88 &  $<$1.00 &  0.54 &  44.447 & 43.552 \\
084504.20$+$542612.0 & 0.30317 & 7.97 & 1.53 &  $<$1.00 &  0.88 &  44.427 & 43.895 \\
084853.09$+$282411.8 & 0.19820 & 8.29 & 0.81 &  $<$1.00 &  0.60 &  44.265 & 43.715 \\
085259.22$+$031320.6 & 0.29708 & 8.31 & 8.80 &  $<$1.00 &$-$0.02&  44.334 & 43.979 \\
085632.39$+$504114.0 & 0.23471 & 8.20 & 5.11 &  $<$1.00 &$-$0.10&  44.005 & 43.129 \\
085828.69$+$342343.8 & 0.25666 & 8.70 & 6.43 &  $<$1.00 &$-$0.08&  45.039 & 43.989 \\
085900.48$+$383211.6 & 0.34551 & 7.73 & 2.94 &  $<$1.00 &  0.65 &  45.459 & 43.663 \\
085915.65$+$011800.5 & 0.28206 & 8.40 & 3.25 &  $<$1.00 &  0.46 &  44.582 & 43.603 \\
090137.99$+$532051.1 & 0.16165 & 7.11 & 1.05 &  $<$1.00 &  0.25 &  44.464 & 43.359 \\
090151.14$+$103020.4 & 0.20093 & 7.92 & 4.13 &  $<$1.00 &$-$0.14&  44.572 & 43.903 \\
090455.00$+$511444.6 & 0.22463 & 8.20 & 2.55 &  $<$1.00 &  0.10 &  44.364 & 43.564 \\
090519.67$+$440139.1 & 0.34246 & 7.50 & 1.64 &  $<$1.00 &  0.80 &  44.898 & 43.344 \\
090654.47$+$391455.4 & 0.24067 & 7.39 & 1.85 &  $<$1.00 &  0.51 &  44.593 & 43.695 \\
090851.25$+$444611.2 & 0.32021 & 8.34 & 3.66 &  $<$1.00 &  0.42 &  44.130 & 43.885 \\
\end{tabular}
\end{center}

\begin{center}
\begin{tabular}{ccccrrcc}
\hline
\hline
\textbf{Object}&\textbf{Z}&\textbf{log($M_{\rm BH}$)}&\textbf{$L_{\rm 5100\AA}$}&\textbf{$F_{\rm 20cm}$}&\textbf{$R_{\rm i}$}&\textbf{log($L_{\rm 0.1-2.4keV}$)}&\textbf{log($L_{\rm 2-10keV}$)} \\
\textbf{SDSS}& & ($\rm M_{\odot}$)& ($10^{44}$~ergs~s$^{-1}$)  & (mJy) & & ($\rm ergs~s^{-1}$) & ($\rm ergs~s^{-1}$) \\
\textbf{(1)}&\textbf{(2)}&\textbf{(3)}&\textbf{(4)}&\textbf{(5)}&\textbf{(6)}&\textbf{(7)}&\textbf{(8)} \\\hline

090932.04$+$503019.6 & 0.26728 & 7.56 & 1.14 &  $<$1.00 &  0.79 &  43.933 & 43.046 \\
091010.00$+$481341.7 & 0.11700 & 8.14 & 0.86 &  $<$1.00 &  0.00 &  44.297 & 43.446 \\
091702.38$-$004417.5 & 0.32231 & 8.71 & 8.21 &  $<$1.00 &  0.15 &  45.337 & 44.643 \\
091755.02$+$053749.7 & 0.34898 & 8.43 & 1.86 &  $<$1.00 &  0.74 &  44.573 & 43.461 \\
091955.35$+$552137.1 & 0.12295 & 8.46 & 2.33 &  $<$1.00 &$-$0.38&  44.338 & 43.412 \\
092309.87$+$453046.4 & 0.29238 & 7.63 & 2.16 &  $<$1.00 &  0.54 &  44.672 & 43.779 \\
092554.44$+$453544.2 & 0.32948 & 7.97 & 3.10 &  $<$1.00 &  0.47 &  44.534 & 43.631 \\
092909.79$+$464424.0 & 0.23996 & 8.64 & 6.15 &  $<$1.00 &$-$0.12&  43.833 & 43.103 \\
092933.58$+$095617.0 & 0.23194 & 7.53 & 0.95 &  $<$1.00 &  0.70 &  44.199 & 43.811 \\
093701.04$+$010543.7 & 0.05054 & 6.92 & 0.30 &  $<$1.00 &$-$0.04&  43.885 & 43.308 \\
093939.69$+$375705.8 & 0.23127 & 7.79 & 2.24 &  $<$1.00 &  0.34 &  43.739 & 42.863 \\
094621.27$+$471131.3 & 0.23049 & 7.67 & 1.37 &  $<$1.00 &  0.50 &  44.233 & 43.890 \\
095048.38$+$392650.5 & 0.20564 & 8.41 & 3.45 &  $<$1.00 &$-$0.16&  44.807 & 44.329 \\
095302.64$+$380145.2 & 0.27291 & 8.45 & 2.60 &  $<$1.00 &  0.45 &  44.740 & 43.853 \\
095823.45$+$065506.5 & 0.34581 & 8.30 & 5.20 &  $<$1.00 &  0.53 &  44.715 & 44.356 \\
095833.95$+$560224.4 & 0.21639 & 6.99 & 0.74 &  0.83    &  0.61 &  44.113 & 42.926 \\
095915.65$+$050355.1 & 0.16230 & 7.88 & 1.42 &  $<$1.00 &  0.10 &  44.040 & 42.836 \\
095931.67$+$504449.0 & 0.14323 & 7.54 & 1.12 &  $<$1.00 &  0.02 &  43.771 & 42.984 \\
100033.88$+$104723.7 & 0.22648 & 7.99 & 2.25 &  2.17    &  0.29 &  44.776 & 43.656 \\
100201.76$+$620816.3 & 0.13379 & 6.98 & 0.35 &  $<$1.00 &  0.49 &  44.093 & 43.173 \\
100420.13$+$051300.4 & 0.16049 & 7.61 & 2.61 &  $<$1.00 &$-$0.12&  43.401 & 42.547 \\
100541.86$+$433240.4 & 0.17843 & 7.79 & 2.67 &  2.81    &  0.47 &  44.844 & 43.917 \\
100627.94$+$603043.6 & 0.21025 & 7.15 & 1.00 &  $<$1.00 &  0.56 &  44.164 & 43.081 \\
100744.54$+$500746.6 & 0.21204 & 7.67 & 2.51 &  $<$1.00 &  0.18 &  44.594 & 43.861 \\
101044.51$+$004331.3 & 0.17757 & 8.83 & 3.15 &  0.77    &$-$0.36&  44.556 & 43.211 \\
101401.86$+$461953.7 & 0.32112 & 8.38 & 5.56 &  $<$1.00 &  0.22 &  44.707 & 43.696 \\
101415.14$+$091839.3 & 0.25224 & 8.62 & 1.18 &  $<$1.00 &  0.58 &  44.500 & 43.999 \\
101437.46$+$440639.1 & 0.20014 & 7.62 & 1.38 &  $<$1.00 &  0.31 &  44.251 & 43.458 \\
101730.97$+$470225.0 & 0.33499 & 7.91 & 3.62 &  $<$1.00 &  0.48 &  44.629 & 43.732 \\
101852.45$+$495800.3 & 0.15481 & 7.00 & 0.41 &  $<$1.00 &  0.63 &  43.931 & 43.436 \\
102309.48$+$082602.1 & 0.34411 & 8.02 & 2.69 &  $<$1.00 &  0.60 &  44.492 & 43.585 \\
102512.85$+$480853.2 & 0.33156 & 8.50 & 2.35 &  $<$1.00 &  0.66 &  44.289 & 43.386 \\
102531.28$+$514034.8 & 0.04488 & 7.05 & 0.29 &  0.60    &$-$0.43&  44.038 & 42.994 \\
102745.84$+$051558.9 & 0.31480 & 7.56 & 3.02 &  $<$1.00 &  0.45 &  44.648 & 43.411 \\
103421.70$+$605318.1 & 0.22775 & 8.07 & 1.97 &  $<$1.00 &  0.29 &  44.158 & 43.292 \\
103457.29$-$010209.0 & 0.32801 & 7.64 & 3.20 &  $<$1.00 &  0.51 &  44.791 & 43.891 \\
104041.50$+$600239.3 & 0.29711 & 7.63 & 2.53 &  $<$1.00 &  0.54 &  44.318 & 43.656 \\
104541.76$+$520235.5 & 0.28393 & 7.87 & 5.73 &  $<$1.00 &  0.16 &  44.855 & 43.895 \\
105007.75$+$113228.6 & 0.13344 & 7.65 & 2.50 &  1.50    &$-$0.10&  44.264 & 43.445 \\
105055.14$+$552723.2 & 0.33196 & 7.86 & 5.20 &  $<$1.00 &  0.33 &  45.165 & 44.059 \\
105118.23$+$605008.2 & 0.27603 & 7.98 & 1.97 &  $<$1.00 &  0.61 &  44.031 & 42.963 \\
105444.70$+$483139.0 & 0.28651 & 8.73 &10.08 &  1.56    &$-$0.00&  45.219 & 44.523 \\
105752.69$+$105037.9 & 0.22033 & 7.23 & 1.20 &  $<$1.00 &  0.48 &  44.458 & 43.585 \\
105830.13$+$601600.3 & 0.14868 & 7.56 & 1.40 &  $<$1.00 &$-$0.11&  44.300 & 43.041 \\
110016.19$+$393524.5 & 0.31265 & 7.71 & 1.60 &  $<$1.00 &  0.88 &  44.998 & 43.979 \\
110540.09$+$064225.7 & 0.23046 & 8.36 & 1.20 &  $<$1.00 &  0.36 &  44.178 & 44.159 \\
111006.95$+$612521.4 & 0.26234 & 7.60 & 1.46 &  $<$1.00 &  0.51 &  44.403 & 43.527 \\
111706.39$+$441333.3 & 0.14382 & 8.73 & 4.02 &  $<$1.00 &$-$0.45&  43.323 & 42.475 \\
111740.48$+$530151.3 & 0.15851 & 7.56 & 0.83 &  $<$1.00 &  0.36 &  43.783 & 43.081 \\
111830.28$+$402554.0 & 0.15457 & 7.67 & 3.16 &  2.32    &  0.06 &  44.670 & 43.826 \\
112114.21$+$032546.8 & 0.15203 & 7.62 & 1.12 &  2.20    &  0.59 &  43.858 & 43.871 \\
112417.79$+$602026.7 & 0.20470 & 7.27 & 1.06 &  $<$1.00 &  0.49 &  44.288 & 43.619 \\
\end{tabular}
\end{center}

\begin{center}
\begin{tabular}{ccccrrcc}
\hline
\hline
\textbf{Object}&\textbf{Z}&\textbf{log($M_{\rm BH}$)}&\textbf{$L_{\rm 5100\AA}$}&\textbf{$F_{\rm 20cm}$}&\textbf{$R_{\rm i}$}&\textbf{log($L_{\rm 0.1-2.4keV}$)}&\textbf{log($L_{\rm 2-10keV}$)} \\
\textbf{SDSS}& & ($\rm M_{\odot}$)& ($10^{44}$~ergs~s$^{-1}$)  & (mJy) & & ($\rm ergs~s^{-1}$) & ($\rm ergs~s^{-1}$) \\
\textbf{(1)}&\textbf{(2)}&\textbf{(3)}&\textbf{(4)}&\textbf{(5)}&\textbf{(6)}&\textbf{(7)}&\textbf{(8)} \\\hline

112439.18$+$420145.0 & 0.22503 & 8.18 & 7.01 &  $<$1.00 &$-$0.30&  44.623 & 43.072 \\
112646.43$-$013417.9 & 0.34095 & 8.32 & 2.30 &  $<$1.00 &  0.80 &  44.001 & 43.866 \\
112941.93$+$512050.6 & 0.23385 & 7.47 & 2.50 &  1.26    &  0.36 &  44.719 & 43.577 \\
113105.04$+$610405.0 & 0.33802 & 7.95 & 3.64 &  $<$1.00 &  0.46 &  44.206 & 43.415 \\
113422.48$+$041127.6 & 0.10800 & 8.19 & 1.04 &  $<$1.00 &$-$0.26&  44.366 & 43.591 \\
113706.84$+$013948.0 & 0.19262 & 8.73 & 2.64 &  $<$1.00 &  0.07 &  44.714 & 44.322 \\
113738.04$+$103930.1 & 0.17454 & 8.06 & 2.55 &  1.65    &  0.20 &  44.405 & 43.707 \\
113908.97$+$591154.8 & 0.06127 & 7.86 & 0.76 &  $<$1.00 &$-$0.48&  43.543 & 43.303 \\
114039.86$+$414545.9 & 0.23833 & 7.87 & 1.37 &  $<$1.00 &  0.54 &  44.647 & 43.770 \\
114105.71$+$024117.0 & 0.09313 & 7.55 & 0.47 &  $<$1.00 &  0.06 &  43.623 & 43.092 \\
114327.21$+$431145.9 & 0.30505 & 8.05 & 3.38 &  $<$1.00 &  0.44 &  44.447 & 43.550 \\
114341.97$-$014434.5 & 0.10522 & 7.52 & 0.82 &  $<$1.00 &  0.04 &  43.280 & 42.444 \\
114408.90$+$424357.5 & 0.27248 & 7.74 & 2.20 &  $<$1.00 &  0.58 &  44.499 & 43.612 \\
114559.00$+$040409.8 & 0.27359 & 7.69 & 1.31 &  $<$1.00 &  0.88 &  43.916 & 43.889 \\
114954.99$+$044812.9 & 0.26951 & 7.53 & 3.02 &  1.76    &  0.62 &  44.474 & 43.529 \\
115105.41$+$445309.3 & 0.34394 & 7.35 & 1.99 &  $<$1.00 &  0.75 &  44.455 & 43.548 \\
115507.61$+$520129.6 & 0.15399 & 7.30 & 1.12 &  2.05    &  0.58 &  43.652 & 42.801 \\
115549.43$+$502117.1 & 0.28443 & 8.05 & 3.08 &  $<$1.00 &  0.43 &  44.386 & 43.495 \\
115558.97$+$593129.3 & 0.24081 & 7.98 & 2.25 &  $<$1.00 &  0.36 &  44.603 & 43.579 \\
115632.24$+$112653.7 & 0.22582 & 7.52 & 1.54 &  $<$1.00 &  0.38 &  44.176 & 43.302 \\
115758.73$-$002220.7 & 0.25984 & 8.42 & 3.25 &  $<$1.00 &  0.14 &  44.034 & 44.029 \\
120118.43$+$060024.1 & 0.33555 & 8.79 & 2.16 &  $<$1.00 &  0.64 &  44.307 & 43.402 \\
120233.08$+$022559.7 & 0.27287 & 8.51 & 2.85 &  $<$1.00 &  0.37 &  44.197 & 43.366 \\
120347.70$+$520749.7 & 0.17760 & 8.50 & 4.92 &  0.68    &$-$0.54&  44.681 & 43.957 \\
121018.35$+$015405.9 & 0.21589 & 8.40 & 2.71 &  $<$1.00 &  0.22 &  43.768 & 43.174 \\
122420.28$+$435157.9 & 0.32172 & 8.44 & 2.39 &  $<$1.00 &  0.69 &  44.520 & 43.880 \\
122549.28$+$472343.8 & 0.31883 & 7.56 & 2.04 &  $<$1.00 &  0.66 &  44.294 & 43.394 \\
123054.12$+$110011.1 & 0.23596 & 8.29 & 3.67 &  $<$1.00 &$-$0.02&  44.503 & 44.301 \\
123115.20$+$590707.2 & 0.32605 & 8.05 & 3.14 &  $<$1.00 &  0.48 &  44.461 & 44.272 \\
123220.11$+$495721.7 & 0.26188 & 7.65 & 2.94 &  $<$1.00 &  0.50 &  45.082 & 43.929 \\
123234.05$+$092815.2 & 0.34665 & 8.33 & 2.35 &  $<$1.00 &  0.74 &  44.614 & 43.812 \\
123958.57$+$490540.1 & 0.23550 & 7.04 & 1.08 &  $<$1.00 &  0.59 &  43.941 & 43.308 \\
124441.41$+$585626.8 & 0.19806 & 7.25 & 0.75 &  $<$1.00 &  0.54 &  44.667 & 43.269 \\
124635.24$+$022208.7 & 0.04818 & 6.75 & 0.31 &  2.23    &  0.25 &  44.057 & 43.159 \\
124931.72$+$523039.2 & 0.16228 & 8.00 & 2.66 &  $<$1.00 &$-$0.14&  44.447 & 43.272 \\
125519.69$+$014412.3 & 0.34318 & 8.92 &15.52 &  1.47    &  0.11 &  45.350 & 44.548 \\
125719.56$+$442935.4 & 0.30025 & 8.25 & 7.68 &  $<$1.00 &  0.00 &  44.818 & 44.084 \\
125824.57$+$540429.8 & 0.34769 & 8.12 & 2.17 &  $<$1.00 &  0.76 &  44.470 & 43.562 \\
130250.51$+$111827.9 & 0.20284 & 7.47 & 1.71 &  $<$1.00 &  0.28 &  44.424 & 43.334 \\
130416.99$+$020537.1 & 0.22854 & 7.67 & 2.04 &  $<$1.00 &  0.27 &  45.378 & 43.966 \\
130421.84$+$560817.0 & 0.29830 & 7.82 & 2.17 &  $<$1.00 &  0.62 &  44.406 & 43.823 \\
130604.48$+$453405.4 & 0.32709 & 7.52 & 2.17 &  $<$1.00 &  0.80 &  44.683 & 44.095 \\
131308.67$+$542115.5 & 0.29798 & 8.45 & 2.90 &  $<$1.00 &  0.63 &  44.455 & 43.872 \\
131404.97$+$153054.2 & 0.16517 & 7.94 & 1.27 &  $<$1.00 &  0.22 &  44.285 & 43.363 \\
131651.70$+$630719.9 & 0.34724 & 8.65 & 2.01 &  $<$1.00 &  0.79 &  44.110 & 44.246 \\
132144.96$+$033055.7 & 0.26889 & 8.70 & 5.48 &  $<$1.00 &  0.02 &  44.386 & 43.765 \\
132242.47$-$022522.0 & 0.12115 & 7.69 & 0.68 &  $<$1.00 &  0.18 &  43.569 & 43.592 \\
132447.65$+$032432.6 & 0.30578 & 8.25 & 4.70 &  $<$1.00 &  0.16 &  44.358 & 43.461 \\
132643.62$+$015209.4 & 0.19670 & 7.68 & 0.85 &  $<$1.00 &  0.52 &  44.339 & 43.680 \\
132802.54$+$441805.3 & 0.23155 & 8.13 & 0.77 &  $<$1.00 &  0.84 &  43.756 & 42.880 \\
133051.24$+$412858.1 & 0.18174 & 7.70 & 1.58 &  $<$1.00 &  0.17 &  44.429 & 43.511 \\
133300.83$+$451809.0 & 0.31974 & 8.36 & 3.65 &  $<$1.00 &  0.37 &  44.727 & 44.267 \\
\end{tabular}
\end{center}

\begin{center}
\begin{tabular}{ccccrrcc}
\hline
\hline
\textbf{Object}&\textbf{Z}&\textbf{log($M_{\rm BH}$)}&\textbf{$L_{\rm 5100\AA}$}&\textbf{$F_{\rm 20cm}$}&\textbf{$R_{\rm i}$}&\textbf{log($L_{\rm 0.1-2.4keV}$)}&\textbf{log($L_{\rm 2-10keV}$)} \\
\textbf{SDSS}& & ($\rm M_{\odot}$)& ($10^{44}$~ergs~s$^{-1}$)  & (mJy) & & ($\rm ergs~s^{-1}$) & ($\rm ergs~s^{-1}$) \\
\textbf{(1)}&\textbf{(2)}&\textbf{(3)}&\textbf{(4)}&\textbf{(5)}&\textbf{(6)}&\textbf{(7)}&\textbf{(8)} \\
\hline

133423.30$+$434331.7 & 0.22551 & 7.65 & 1.31 &  $<$1.00 &  0.58 &  44.290 & 43.751 \\
133636.65$+$420934.1 & 0.22328 & 8.23 & 2.46 &  3.55    &  0.73 &  43.559 & 42.685 \\
134032.02$+$052158.5 & 0.27448 & 8.73 & 2.26 &  $<$1.00 &  0.63 &  44.668 & 43.892 \\
134206.57$+$050523.8 & 0.26602 & 7.59 & 3.48 &  3.85    &  0.90 &  44.468 & 43.582 \\
134251.60$-$005345.2 & 0.32581 & 8.37 & 6.22 &  $<$1.00 &  0.26 &  44.366 & 43.364 \\
134444.17$+$630337.2 & 0.29218 & 7.80 & 0.98 &  $<$1.00 &  0.85 &  44.024 & 43.327 \\
134459.45$-$001559.5 & 0.24490 & 7.55 & 1.41 &  $<$1.00 &  0.42 &  44.472 & 43.475 \\
134845.44$+$451809.5 & 0.27653 & 8.90 & 3.42 &  $<$1.00 &  0.34 &  44.715 & 43.930 \\
135829.58$+$010908.4 & 0.24392 & 8.30 & 1.76 &  $<$1.00 &  0.60 &  44.118 & 43.240 \\
135946.91$+$581357.0 & 0.22391 & 7.44 & 0.68 &  $<$1.00 &  0.70 &  44.096 & 43.716 \\
140050.21$+$532424.5 & 0.17466 & 7.25 & 1.06 &  $<$1.00 &  0.27 &  43.538 & 42.678 \\
140104.18$+$635234.0 & 0.34386 & 8.44 & 3.47 &  $<$1.00 &  0.48 &  44.223 & 42.802 \\
140604.25$+$572956.5 & 0.32581 & 8.57 & 2.44 &  $<$1.00 &  0.67 &  43.943 & 43.040 \\
140824.76$+$543817.7 & 0.29106 & 7.63 & 2.36 &  $<$1.00 &  0.59 &  44.441 & 43.548 \\
140827.51$+$142233.2 & 0.31766 & 7.48 & 1.62 &  $<$1.00 &  0.84 &  44.628 & 43.795 \\
141046.46$+$465802.9 & 0.33409 & 7.38 & 2.99 &  $<$1.00 &  0.65 &  44.873 & 43.164 \\
141758.25$+$360741.4 & 0.21180 & 7.62 & 0.81 &  $<$1.00 &  0.44 &  44.273 & 43.558 \\
141942.46$+$591259.4 & 0.31988 & 8.37 & 1.59 &  $<$1.00 &  0.89 &  44.012 & 43.318 \\
142424.22$+$595300.5 & 0.13495 & 8.75 & 2.28 &  2.51    &  0.08 &  44.091 & 43.537 \\
142455.53$+$421407.6 & 0.31603 & 8.19 & 7.32 &  $<$1.00 &  0.17 &  44.661 & 43.361 \\
142734.80$+$352543.4 & 0.34038 & 7.49 & 1.69 &  $<$1.00 &  0.78 &  44.084 & 43.177 \\
142748.29$+$050222.0 & 0.10608 & 7.24 & 1.27 &  $<$1.00 &$-$0.26&  43.657 & 42.900\\
143039.30$+$493538.9 & 0.20355 & 7.82 & 1.86 &  $<$1.00 &  0.32 &  44.544 & 43.403 \\
143204.60$+$394439.0 & 0.34848 & 9.05 & 9.25 &  $<$1.00 &  0.14 &  44.063 & 43.154 \\
143704.12$+$000705.0 & 0.14036 & 7.45 & 0.61 &  $<$1.00 &  0.44 &  44.165 & 43.270 \\
143919.31$+$551317.8 & 0.25720 & 8.52 & 1.10 &  $<$1.00 &  0.77 &  44.245 & 43.807 \\
143940.27$+$030528.6 & 0.26856 & 7.23 & 1.75 &  $<$1.00 &  0.64 &  45.053 & 43.949 \\
144012.76$+$615633.0 & 0.27547 & 7.54 & 4.46 &  2.88    &  0.69 &  44.860 & 43.701 \\
144050.76$+$520445.9 & 0.31834 & 7.89 & 3.73 &  $<$1.00 &  0.46 &  44.684 & 43.861 \\
144202.82$+$433708.7 & 0.23146 & 8.18 & 3.54 &  $<$1.00 &  0.20 &  44.584 & 44.196 \\
144302.59$+$404525.1 & 0.24615 & 7.79 & 3.16 &  $<$1.00 &  0.11 &  44.231 & 43.673 \\
144645.93$+$403505.7 & 0.26709 & 8.35 &11.37 &  $<$1.00 &$-$0.13&  44.999 & 44.073\\
144751.79$+$505328.7 & 0.30565 & 7.56 & 1.41 &  $<$1.00 &  0.86 &  44.068 & 43.173 \\
145006.93$+$581456.9 & 0.31520 & 7.72 & 3.93 &  $<$1.00 &  0.33 &  44.714 & 43.755 \\
145254.74$+$395714.6 & 0.22528 & 7.31 & 0.79 &  $<$1.00 &  0.72 &  43.315 & 43.261 \\
145307.63$+$380319.6 & 0.26191 & 7.32 & 1.51 &  $<$1.00 &  0.61 &  44.695 & 43.416 \\
145439.39$+$465309.1 & 0.31645 & 8.45 & 1.95 &  $<$1.00 &  0.58 &  44.502 & 44.050 \\
145440.39$+$455008.5 & 0.32471 & 7.92 & 1.24 &  $<$1.00 &  0.86 &  44.017 & 43.115 \\
145817.42$+$455514.8 & 0.28575 & 7.79 & 3.12 &  $<$1.00 &  0.41 &  44.669 & 43.422 \\
145824.46$+$363119.5 & 0.24637 & 7.91 & 3.86 &  $<$1.00 &  0.20 &  44.384 & 43.474 \\
150049.15$+$343727.1 & 0.31794 & 7.52 & 1.95 &  $<$1.00 &  0.74 &  44.423 & 43.423 \\
150114.45$+$393927.7 & 0.23072 & 7.51 & 1.68 &  $<$1.00 &  0.49 &  43.950 & 43.074 \\
150155.24$+$563731.7 & 0.34338 & 7.54 & 2.78 &  $<$1.00 &  0.56 &  44.370 & 42.992 \\
150242.39$+$395208.5 & 0.34185 & 8.19 & 6.59 &  $<$1.00 &  0.22 &  44.894 & 43.976 \\
150437.67$+$541149.6 & 0.30519 & 8.39 & 3.17 &  $<$1.00 &  0.45 &  44.154 & 43.128 \\
151304.56$+$304204.2 & 0.34801 & 8.14 & 5.25 &  $<$1.00 &  0.25 &  44.886 & 43.997 \\
151557.75$+$302357.1 & 0.25171 & 7.60 & 0.94 &  $<$1.00 &  0.70 &  44.117 & 43.235 \\
151940.87$+$320157.9 & 0.30132 & 8.35 & 2.40 &  $<$1.00 &  0.58 &  44.558 & 43.896 \\
152153.81$+$594020.0 & 0.28599 & 8.10 & 2.41 &  $<$1.00 &  0.52 &  44.260 & 44.001 \\
152524.16$+$523205.7 & 0.32755 & 8.28 & 2.85 &  $<$1.00 &  0.59 &  44.107 & 43.125 \\
152628.19$-$003809.4 & 0.12333 & 7.40 & 0.57 &  $<$1.00 &  0.46 &  43.696 & 42.856 \\
153941.49$+$504255.8 & 0.20315 & 7.40 & 2.32 &  $<$1.00 &  0.14 &  44.766 & 43.774 \\
\end{tabular}
\end{center}

\begin{center}
\begin{tabular}{ccccrrcc}
\hline
\hline
\textbf{Object}&\textbf{Z}&\textbf{log($M_{\rm BH}$)}&\textbf{$L_{\rm 5100\AA}$}&\textbf{$F_{\rm 20cm}$}&\textbf{$R_{\rm i}$}&\textbf{log($L_{\rm 0.1-2.4keV}$)}&\textbf{log($L_{\rm 2-10keV}$)} \\
\textbf{SDSS}& & ($\rm M_{\odot}$)& ($10^{44}$~ergs~s$^{-1}$)  & (mJy) & & ($\rm ergs~s^{-1}$) & ($\rm ergs~s^{-1}$) \\
\textbf{(1)}&\textbf{(2)}&\textbf{(3)}&\textbf{(4)}&\textbf{(5)}&\textbf{(6)}&\textbf{(7)}&\textbf{(8)} \\\hline

154004.24$+$355050.1 & 0.16361 & 8.16 & 1.31 &  $<$1.00 &  0.17 &  43.955 & 43.126 \\
154344.27$-$001452.1 & 0.30189 & 8.16 & 3.01 &  $<$1.00 &  0.55 &  44.512 & 43.620 \\
154348.62$+$401324.9 & 0.31826 & 8.60 & 4.93 &  2.75    &  0.84 &  44.885 & 44.080 \\
155207.18$+$525347.0 & 0.33519 & 8.22 & 2.50 &  $<$1.00 &  0.65 &  44.313 & 43.891 \\
155324.25$+$490726.9 & 0.25807 & 8.19 & 1.88 &  $<$1.00 &  0.59 &  44.297 & 43.414 \\
155328.49$+$095102.0 & 0.19174 & 8.83 & 3.31 &  $<$1.00 &$-$0.04&  43.859 & 42.996 \\
155837.88$+$282839.0 & 0.32336 & 8.54 & 8.80 &  $<$1.00 &  0.12 &  45.104 & 44.163 \\
160518.50$+$375653.5 & 0.20087 & 7.55 & 1.64 &  $<$1.00 &  0.22 &  44.229 & 43.314 \\
161118.82$+$291932.4 & 0.29364 & 8.29 & 1.68 &  $<$1.00 &  0.74 &  44.127 & 43.757 \\
161849.25$+$442517.2 & 0.33521 & 8.46 & 7.97 &  3.09    &  0.63 &  44.636 & 44.047 \\
162607.24$+$335915.2 & 0.20453 & 8.10 & 5.54 &  3.74    &  0.44 &  44.435 & 43.243 \\
163111.28$+$404805.2 & 0.25751 & 8.03 & 2.18 &  $<$1.00 &  0.29 &  44.789 & 44.152 \\
163631.58$+$461704.3 & 0.25170 & 7.64 & 1.55 &  $<$1.00 &  0.52 &  43.988 & 43.086 \\
164343.24$+$405654.3 & 0.34375 & 7.04 & 1.34 &  $<$1.00 &  0.95 &  43.997 & 43.089 \\
165338.69$+$634010.6 & 0.27897 & 7.54 & 1.91 &  $<$1.00 &  0.60 &  44.215 & 41.701 \\
165352.82$+$384542.1 & 0.32302 & 7.64 & 1.74 &  $<$1.00 &  0.80 &  44.414 & 43.383 \\
165737.30$+$604939.3 & 0.31316 & 7.67 & 1.48 &  $<$1.00 &  0.92 &  43.723 & 43.004 \\
170231.06$+$324719.6 & 0.16333 & 7.76 & 4.13 &  1.52    &$-$0.14&  44.858 & 43.879 \\
170302.88$+$191033.9 & 0.29045 & 7.86 & 4.44 &  $<$1.00 &  0.38 &  44.980 & 44.210 \\
170525.54$+$194722.7 & 0.19392 & 8.42 & 1.33 &  $<$1.00 &  0.48 &  44.246 & 43.436 \\
171013.42$+$334402.6 & 0.20749 & 8.67 & 5.73 &  4.60    &  0.40 &  44.590 & 43.730 \\
171207.44$+$584754.5 & 0.26925 & 7.69 & 1.55 &  $<$1.00 &  0.64 &  44.551 & 43.446 \\
171601.93$+$311213.8 & 0.11015 & 7.57 & 1.76 &  2.42    &$-$0.06&  44.301 & 43.686 \\
171750.59$+$581514.0 & 0.31014 & 8.37 & 4.19 &  $<$1.00 &  0.42 &  44.428 & 43.708 \\
171850.30$+$304201.6 & 0.28176 & 7.39 & 2.03 &  0.64    &  0.43 &  44.500 & 43.400 \\
171902.29$+$593715.8 & 0.17852 & 7.49 & 1.06 &  $<$1.00 &  0.29 &  44.129 & 43.018 \\
172711.81$+$632241.8 & 0.21736 & 9.17 & 2.71 &  1.23    &  0.23 &  44.482 & 43.647 \\
173229.44$+$564811.3 & 0.30320 & 7.62 & 1.87 &  $<$1.00 &  0.67 &  44.309 & 43.555 \\
211204.85$-$063535.2 & 0.20419 & 7.75 & 1.71 &  $<$1.00 &  0.34 &  44.239 & 43.890 \\
213818.97$+$011222.4 & 0.34409 & 7.99 & 3.98 &  $<$1.00 &  0.48 &  44.584 & 44.853 \\
215010.52$-$001000.7 & 0.33481 & 8.41 & 1.90 &  $<$1.00 &  0.80 &  44.719 & 43.833 \\
215516.14$+$003250.8 & 0.27782 & 7.62 & 2.11 &  $<$1.00 &  0.59 &  44.648 & 43.498 \\
231250.88$+$001719.0 & 0.25704 & 7.56 & 2.02 &  $<$1.00 &  0.56 &  44.201 & 43.319 \\
233512.68$-$100040.3 & 0.24288 & 7.92 & 0.50 &  $<$1.00 &  0.83 &  44.095 & 43.977 \\
\hline
\end{tabular}
\end{center}
\end{tiny}
 \emph{Notes.}------\\
 Col. 1, object name in J2000.0. Col. 2, redshift
given by the SDSS spectroscopic pipeline. Col. 3, logarithm of black
hole mass . Col. 4, luminosity at $5100\AA$ in unit of $10^{44}$~ergs~s$^{-1}$. Col. 5, flux density at
20cm, '$<$1' represents that the object is covered yet not detected
by FIRST. Col. 6, the radio--loudness. Col. 7--8, X-ray luminosity
of 0.1--2.4 keV and  2--10 keV. \\

%

\label{lastpage}

\begin{thebibliography}{99}

\bibitem[2005]{bian05} Bian W. H., 2005, ChJAS, 5, 289 
 \bibitem[2007]{bon07} Bonning E.W., Cheng L. et al., 2007, ApJ, 659, 211
    \bibitem[]{} Cao Xinwu, 2009, MNRAS, 394, 207
    \bibitem[2008]{dong08} Dong X.B., Wang T.G. et al., 2008, MNRAS, 383, 581(Dong08)
  \bibitem[1994]{elvis94} Elvis M., et al., 1994, ApJS, 95, 1
    \bibitem[1979]{gal79} Galeev A. A.,Rosner R., Vaiana G. S.,1979, ApJ, 229, 318
  \bibitem[]{} Haardt, F., \& Maraschi, L., 1991, ApJ, 380, L51	
  \bibitem[]{} Haardt, F., \& Maraschi, L., 1993, ApJ, 413, 507 
\bibitem[2002]{Ive02} Ivezi\'{c} \v{Z}., et al., 2002, AJ, 124, 2364
   \bibitem[]{} Kawaguchi, T., Shimura, T., \& Mineshige, S., 2001, ApJ, 546, 966 
    \bibitem[1977]{lp77} Liang E. P. T., Price R. H., 1977, ApJ, 218, 247
  \bibitem[1999]{liu99} Liu B. F., Yuan W., Meyer F., et al., 1999, ApJ, 527, L17
  \bibitem[]{} Liu B. F., Mineshige S., Meyer F., et al. 2002, ApJ, 575, 117 (2002a)
  \bibitem[]{} Liu B. F., Mineshige S., Shibata K., 2002, ApJ, 572, L173 (2002b)
\bibitem[2003]{liu03} Liu B. F., Mineshige S., Ohsuga K., 2003, ApJ, 587, 571
  \bibitem[2007]{liu07} Liu B. F., Taam Ronald E., Meyer-Hofmeister E., Meyer F., 2007, ApJ, 671, 695
      \bibitem[1999]{luyu99} Lu Y. \& Yu Q., 1999, ApJ, 570, L47
\bibitem[1974]{lp74} Lynden--Bell D. \& Pringle J.E., 1974, MNRAS, 168, 603
    \bibitem[1994]{mm94} Meyer F. \& Meyer--Hofmeister E., 1994, A\&A, 288, 175
  \bibitem[2000]{me200} Meyer F., Liu B. F., et al.,2000, A\&A, 361, 175
   \bibitem[2008]{mid08} Middleton M., Done C. and Schurch N., 2008, MNRAS, 383, 1501
\bibitem[]{} Nakamura, K \& Osaki, Y., 1993, PASJ, 45, 775
  \bibitem[1994]{ny94} Narayan R. \& Yi I., 1994, ApJ, 428, L13
  \bibitem[1998]{na98} Narayan R., Mahadevan R., \& Quataert E., 1998, in The~Thoery~of~Black~Hole~Accretion~Discs, eds. M. A. Abramowicz, G. Bjornsson, \& J. E. Pringle (Cambridge University, Cambridge)
   \bibitem[]{} Narayan R., \& Yi, I. 1995, ApJ, 452, 710  
\bibitem[1993]{peterson93} Peterson B. M., 1 Narayan et al.993, PASP, 100, 18
  \bibitem[2004]{peterson04} Peterson B. M., Ferrarese L., et al., 2004, ApJ, 613, 682
  \bibitem[1990]{pounds90} Pounds K. A., Nandran K., Stewart G. C. et al., 1990, Nature, 344, 132
  \bibitem[2000]{retu00} Reeves J. N., Turner M. J. L., 2000, MNRAS, 316, 234
\bibitem[]{} Saez C., Chartas G., et al. 2008, ApJ, 135, 1505
\bibitem[1973]{SS73} Shakura N.I., Sunyaev R.A., 1973, A\&A, 24, 337
   \bibitem[1984]{stell84} Stella L., Rosner R., 1984, ApJ, 227, 312
\bibitem[]{} Svensson R., Zdziarski A. A., 1994, ApJ, 436, 599
\bibitem[]{}Taam, R.E, Liu, B. F, et al., 2008, ApJ, 688, 527  
\bibitem[1974]{thone74} Thorne K. S., 1974, ApJ, 191, 507
  \bibitem[]{}Vasudevan R .V., Fibian A. C., 2007, MNRAS,381, 1235
  \bibitem[2006]{vest06} Vestergaard M., \& Peterson B. M., 2006, ApJ, 641, 689
\bibitem[1994]{voges94} Voges W., Gruber R., Haberl F., et al., 1994, ROSAT NEWS No.32
\bibitem[2004]{wang04} Wang J. M., Watarai K. Y., Mineshige, S., 2004, ApJ, 607, L107
    \bibitem[2000]{will00} William H.P., Saul A.T. et al., 2000, Numerical~Recipes~in~Fortran (
 2nd, Cambridge University Press, Cambridge)
  \bibitem[2002]{woo02} Woo J. H. \& Urry C. M., 2002, ApJ, 578, 530
  \bibitem[2007]{yang07} Yang F., Hu C. et al., 2007, ChJAA, 7, 353
\bibitem[]{} Zhou Y. Y., Yu K. N., et al., 1997, ApJ, L9



 \end{thebibliography}
\end{document}